\documentstyle[prl,aps]{revtex}
\evensidemargin=\oddsidemargin
\begin{document}
\draft
\title{$\bf {S-D}$ mixing and searching for the $\bf\psi(1^1P_1)$ state at
the Beijing Electron-Positron Collider}
\bigskip
\author{Yu-Ping Kuang}
\address{China Center of Advanced Science and Technology (World
Laboratory), P.O.Box 8730, Beijing 100080, China\\
and Department of Physics, Tsinghua University, Beijing
100084, China.\footnote{Mailing address.}}
\bigskip\bigskip
\date{TUHEP-TH-01129}
\maketitle
\begin{abstract}

The $\psi(1^1P_1)$ state can be produced at the Beijing
Electron-Positron Collider (BEPC) in the process 
$~\psi^\prime\to \psi(1^1P_1)+\pi^0~$. We calculate the rate of this
process taking account of the $S-D$ mixing effect in $\psi^\prime$. It
is shown that the rate is about a factor of 3 smaller than the simple
result without considering the $S-D$ mixing effect. Possible detecting
channels are suggested and it is shown that $\psi(1^1P_1)$ is able to be found
with the accumulation of $3\times 10^7$ events of $\psi^\prime$ at BEPC.
\end{abstract}

\vspace{0.2cm}\hspace{1.84cm}
PACS number(s): 13.20.Gd, 13.30.Eg, 14.40.Cs
\bigskip\bigskip

Searching for the $^1P_1$ states of heavy quarkonia is of interest
since the difference between the $^1P_1$ mass $M_{^1P_1}$ and the
center-of-gravity (c.o.g) of the $^3P_J$ mass 
$M_{c.o.g}=(5M_{^3P_2}+3M_{^3P_1}+M_{^3P_0})/9$ gives useful information about 
the spin-dependent interactions between the heavy quark and 
antiquark. Theoretical investigations of the general structure of the
spin-dependent interactions and the formula for the spin-dependent
potential up to $O(1/m^2)$ in the $1/m$ expansion have been carried out in
various approaches \cite{SDI,SDP}. Experimentally, neither the
$\psi(1^1P_1)$ state nor the $\Upsilon(1^1P_1)$ state has been found
yet\footnote{In 1992, the E760 collaboration claimed that they saw a
significant enhancement in $\bar{p}+p\to J/\psi+\pi^0$ at  
$\sqrt{s}= 3526.2$ MeV which was supposed to be a candidate of
$\psi(1^1P_1)$ \cite{E760}. However, such an enhancement has not been 
confirmed by the successive E835 experiment with significantly higher 
statistics \cite{E835}.}. The best way of searching for the $\psi(1^1P_1)$ state
is to look for the process
\begin{eqnarray}                    
\psi^\prime\to \psi(1^1P_1)+\pi^0
\label{1P1prod}
\end{eqnarray}
at an $e^+e^-$ collider in the $\tau$-charm energy range. Many years
ago, the Crystal Ball Group searched for this process with a negative
result, and the obtained upper limit for the branching ratio is \cite{CB}
\begin{eqnarray}                     
B(\psi^\prime\to \psi(1^1P_1)\pi^0)<0.46\%,~~~~~~95\%~{\rm C.L.}
\label{CBlimit}
\end{eqnarray}
for $M_{\psi(1^1P_1)}=3520$ MeV. Now the most promising collider for this
kind of experiment is Beijing Electron-Positron Collider (BEPC).

Theoretically, a simple single-channel calculation of the transition
rate (\ref{1P1prod}) based on the QCD multipole expansion \cite{Yan,KY81} using 
the Cornell potential model \cite{Cornell} was given in our previous paper
\cite{KTY}. The obtained results are
\begin{eqnarray}                      
&&\Gamma(\psi^\prime\to
\psi(1^1P_1)\pi^0)=0.12\bigg(\frac{\alpha_M}{\alpha_E}\bigg)~{\rm keV},
\nonumber\\
&&B(\psi^\prime\to \psi(1^1P_1)\pi^0)=(4.3\pm 0.5)
\bigg(\frac{\alpha_M}{\alpha_E}\bigg)
\times 10^{-4},
\label{KTY}
\end{eqnarray}
where $\alpha_E=g^2_E/4\pi$ and $\alpha_M=g^2_M/4\pi$ are phenomenological 
coupling constants for color electric dipole and color magnetic dipole gluon 
radiations, respectively. The branching ratio here is obtained from the updated
total width $\Gamma_{tot}(\psi^\prime)=227\pm 31$ keV \cite{PDG}, and it is a 
little different from the value in Ref.\cite{KTY}. The uncertainty in the 
branching ratio comes from the experimental error in 
$\Gamma_{tot}(\psi^\prime)$. 
Phenomenological determination of the ratio $\alpha_M/\alpha_E$ is not so 
certain \cite{KTY}. From the theoretical point of view, the two coupling 
constants should not be so different, so we roughly take a possible range
\begin{eqnarray}                      
\frac{\alpha_M}{\alpha_E}\approx 1\--3.
\label{M/E}
\end{eqnarray}
in the calculation. We see that the theoretically predicted $B(\psi^\prime\to
\psi(1^1P_1)\pi^0)$ is consistent with the Crystal Ball limit
(\ref{CBlimit}), and is not much smaller than the limit, thus it is
hopeful that we will find the $\psi(1^1P_1)$ state in the near future at BEPC.

Since the $\psi^\prime$ mass $M_{\psi^\prime}=3685.96\pm 0.09$ MeV is quite 
close to the $D\bar{D}$ threshold $E_{th}=3729.0\pm 1.0$ MeV \cite{PDG}, it is
expected that the coupled-channel effect may affect the transition rate.
Complete coupled-channel calculation of the transition rate is
tedious.
A coupled-channel formulation of the hadronic transitions between heavy 
quarkonium states and the transition rates for $\Upsilon^\prime\to 
\Upsilon\pi\pi,~\Upsilon^{\prime\prime}\to \Upsilon\pi\pi$, and 
$\Upsilon^{\prime\prime}\to \Upsilon^\prime\pi\pi$ have been given in 
Ref.\cite{ZK}.
Further improved study of the coupled-channel theory of hadronic transitions 
in the heavy quarkonium systems will be presented in a future paper. 
We notice that an important aspect of the
coupled-channel effect affecting the transition rate is the state mixing.
From a reasonable coupled-channel model, the unitarized quark model
(UQM) \cite{UQM}, we see that $\psi^\prime$ is not a pure $2S$ state.
Apart from negligibly small ingredients (an order of magnitude smaller),
$\psi^\prime$ mainly contains $\psi_{2S}$ and $\psi_{1D}$ states. We denote
this mixing as
\begin{eqnarray}                     
&&\psi^\prime=\psi_{2S}\cos\theta+\psi_{1D}\sin\theta,\nonumber\\
&&\psi^{\prime\prime}=-\psi_{2S}\sin\theta+\psi_{1D}\cos\theta.
\label{mixing}
\end{eqnarray}
The UQM gives $\theta\approx -8.5^\circ$ \cite{UQM}.
It is the purpose of this paper to calculate the influence of such an $S-D$ 
mixing on the transition rate (\ref{1P1prod})\footnote{From Ref.\cite{UQM}, we 
see that state-mixings in $\Upsilon^{\prime\prime}$ is an order of
magnitude smaller, so we expect that the single-channel result of the
rate of $\Upsilon^{\prime\prime}\to \Upsilon(1^1P_1)\pi\pi$ given in 
Ref.\cite{KTY} will not be much affected by the state-mixing effect.}.

Instead of taking a specific model of the coupled-channel effect, we take a
simple and phenomenological approach such as was done in studying
hadronic transitions of $\psi(3770)$ in Ref.\cite{KY90} and in studying
radiative transitions of heavy quarkonia in Ref.\cite{Godfrey},
i.e., we determine the mixing angle $\theta$ by fitting the experimental value 
of the ratio of the leptonic decay rates $\Gamma(\psi^\prime\to
e^+e^-)/\Gamma(\psi^{\prime\prime}\to e^+e^-)$. In the simple Cornell 
potential model, the so-determined $\theta$ is \cite{KY90}
\begin{eqnarray}                       
\theta=-10^\circ.
\label{thetaCornell}
\end{eqnarray}
This is consistent with the mixing angle $\theta\approx -8.5^\circ$ in the UQM 
mentioned above. In this paper, we take an improved potential model by Chen 
and Kuang, which reflects more about QCD and leads to more successful 
phenomenological results \cite{CK}. The potential reads
\begin{eqnarray}                      
\displaystyle
V(r)=kr-\frac{16\pi}{25}\frac{1}{rf(r)}\left[1+\frac{2\gamma_E+
\displaystyle\frac{53}{75}}{f(r)}-\frac{462}{625}\frac{\ln f(r)}{f(r)}\right],
\label{CKpot}
\end{eqnarray}
where $k=0.1491~{\rm GeV}^2$ is the string tension related to the Regge slope 
\cite{CK}, $\gamma_E$ is the Euler constant, and $f(r)$ is
\begin{eqnarray}                      
\displaystyle
f(r)=\ln\left[\frac{1}{\Lambda_{\overline{MS}}~r}+4.62-\bigg(1-\frac{1}{4}
\frac{\Lambda_{\overline{MS}}}{\Lambda^I_{\overline{MS}}}\bigg)
\frac{1-\exp\bigg\{-\bigg[15\bigg(3\displaystyle
\frac{\Lambda^I_{\overline{MS}}}{\Lambda_{\overline{MS}}}
-1\bigg)\Lambda_{\overline{MS}}~r\bigg]^2\bigg\}}
{\Lambda_{\overline{MS}}~r}\right]^2,
\label{f(r)}
\end{eqnarray}
in which $\Lambda^I_{\overline{MS}}=180$ MeV. Taking
$\Lambda_{\overline{MS}}=200$ MeV \footnote{Our calculation shows that the
determined $\theta$ is not sensitive to the value of
$\Lambda_{\overline{MS}}$. For example, a 50 MeV variation of
$\Lambda_{\overline{MS}}$ causes only a  $<1\%$ change of $\theta$.},
we obtain the following wave functions at the origin
\begin{eqnarray}                       
\psi_{2S}(0)=0.199~{\rm GeV}^{3/2},\hspace{3cm}
\frac{5}{\sqrt{2}}\frac{\psi^{\prime\prime}_{1D}(0)}{2m_c^2}
=0.0262~{\rm GeV}^{3/2},
\label{psi(0)}
\end{eqnarray}
and then, as was done in Ref.\cite{KY90}, we determine
\begin{eqnarray}                    
\theta=-12^\circ
\label{thetaCK}
\end{eqnarray}
in this model. This is close to the value (\ref{thetaCornell}) in the Cornell 
model.

In general, the initial-state quarkonium $\Phi_i$ and the final-state 
quarkonium $\Phi_f$ in the hadronic transition $\Phi_i\to \Phi_f + h$
[$h$ stands for light hadron(s)] can be described as
\begin{eqnarray}                     
\Phi_i=\sum_{n_il_i}a^{(i)}_{n_il_i}\phi_{n_il_i},~~~~~~
\Phi_f=\sum_{n_fl_f}a^{(f)}_{n_fl_f}\phi_{n_fl_f},
\label{Phi}
\end{eqnarray}
where $\phi_{n_il_i}$ ($\phi_{n_fl_f}$) is the  quarkomium state with 
principal quantum number $n_i(n_f)$ and orbital angular momentum $l_i(l_f)$,
and the mixing coefficients satisfy $\sum_{n_il_i}a^{(i)}_{n_il_i}=1$ and 
$\sum_{n_fl_f}a^{(f)}_{n_fl_f}=1$. In the framework
of the QCD multipole expansion \cite{Yan,KY81}, the hadronic transition 
(\ref{1P1prod}) is mainly contributed by the $E_1M_1$ transition, and the 
transition amplitude is
\begin{eqnarray}                      
\displaystyle
{\cal M}_{E1M1}=i\frac{g_Eg_M}{6m_Q}\sum_{KL}\left(
\frac{\langle\Phi_f|(s_Q-s_{\bar{Q}})_\alpha|KL\rangle
\langle KL|x_\beta|\Phi_i\rangle}{E_i-E_{KL}}+\frac{\langle\Phi_f|x_\beta
|KL\rangle\langle KL|(s_Q-s_{\bar{Q}})_\alpha|\Phi_i\rangle}{E_i-E_{KL}}\right)
\langle \pi^0|E^a_\alpha B^a_\beta|0\rangle,
\label{E1M1}
\end{eqnarray}
where $m_Q$ is the heavy quark mass, $s_Q(s_{\bar{Q}})$ is the spin of the 
heavy quark (antiquark), $K,~L$ are the principal quantum number and the 
orbital angular momentum of the intermediate state, and $E_i$ and $E_{KL}$ are 
the energy eigenvalues of the initial state and the intermediate state, 
respectively. 

For the process (\ref{1P1prod}), the initial-state quarkonium
$\Phi_i=\psi^\prime$, i.e., $a^{(i)}_{20}=\cos\theta,~a^{(i)}_{12}=\sin\theta$, 
and other $a^{(i)}_{n_il_i}$ vanish. Since the mass of the final-state 
quarkonium $\Phi_f=\psi(1^1P_1)$ is supposed to be close to $M_{c.o.g}=3525.3$ 
MeV \cite{PDG} which is not close to $E_{th}=3729.0\pm 1.0$ MeV, we 
expect that the state mixing effect in $\psi(1^1P_1)$ is small. So that we take
$a^{(f)}_{11}=1$ and other $a^{(f)}_{n_fl_f}$ vanish. The matrix element
$\langle \pi^0|E^a_\alpha B^a_\beta|0\rangle$ can be evaluated by using
the Gross-Treiman-Wilczek formula \cite{GTW}
\begin{eqnarray}                     
\langle\pi^0|g_E^2E^a_\alpha
B^a_\beta|0\rangle
=\frac{1}{3}\delta_{\alpha\beta}
\langle\pi^0|g_E^2{\bf E}^a\cdot {\bf B}^a|0\rangle
=\frac{1}{3}\delta_{\alpha\beta}\frac{4\pi^2}{\sqrt{2}}
\bigg[\frac{m_d-m_u}{m_d+m_u}\bigg]f_\pi m^2_\pi,
\label{GTW}
\end{eqnarray}
where we have identified $g_E$ as the QCD coupling constant $g_s$ as
in Ref.\cite{KTY}. After evaluating the spin- and angular-momentum dependent 
parts of ${\cal M}_{E1M1}$, we obtain
\begin{eqnarray}                      
{\cal M}_{E1M1}=i\frac{g_M}{g_E}\frac{4\pi^2}{18\sqrt{6}m_c}
\left\{\cos\theta\bigg(f^{110}_{2011}+f^{001}_{2011}\bigg)
-\sqrt{2}\sin\theta\bigg(f^{110}_{1211}+f^{201}_{1211}\bigg)\right\}
\bigg[\frac{m_d-m_u}{m_d+m_u}\bigg]f_\pi m^2_\pi,
\label{M(E1M1)}
\end{eqnarray}
where the overlapping integral $f^{LP_iP_f}_{n_il_in_fl_f}$ is defined by
\begin{eqnarray}                        
f^{LP_iP_f}_{n_il_in_fl_f}=\sum_K\frac{\langle R_{n_fl_f}|r^{P_f}|R_{KL}
\rangle\langle R_{KL}|r^{P_i}|R_{n_il_i}\rangle}{E_i-E_{KL}},
\label{f}
\end{eqnarray}
in which $R_{n_il_i},~R_{n_fl_f}$, and $R_{KL}$ are radial wave
functions of the initial-, final-, and intermediate-state quarkonium, 
respectively. The transition rate is then
\begin{eqnarray}                           
\Gamma(\psi^\prime\to \psi(1^1P_1)+\pi^0)
&=&\frac{\pi^3}{143m_c^2}\frac{\alpha_M}{\alpha_E}
\left|\cos\theta\bigg(f^{110}_{2011}+f^{001}_{2011}\bigg)
-\sqrt{2}\sin\theta\bigg(f^{110}_{1211}+f^{201}_{1211}\bigg)\right|^2
\nonumber\\
&&\times\frac{E_{\psi(1^1P_1)}}{M_{\psi^\prime}}\bigg[\frac{m_d-m_u}{m_d+m_u}f_\pi 
m^2_\pi\bigg]^2|{\bf p}_\pi|,
\label{Gamma}
\end{eqnarray}
where
$E_{\psi(1^1P_1)}=(M^2_{\psi^\prime}+M^2_{\psi(1^1P_1)}-m^2_\pi)/
(2M_{\psi^\prime})$ is the energy of $\psi(1^1P_1)$, and $|{\bf p}_\pi|$ is 
the absolute value of the pion momentum.

 As in Ref.\cite{KY81}, when evaluating the transition
amplitude, we describe the intermediate states by the string vibrational 
states proposed in Ref.\cite{BT}. Then for a given potential model, the
radial wave functions and the overlapping integrals 
$f^{LP_iP_f}_{n_il_in_fl_f}$ can be calculated numerically. The results
in the Chen-Kuang potential model with $\Lambda_{\overline{MS}}=200$ MeV are
\begin{eqnarray}                  
\displaystyle
&&\theta=0^\circ:~~~~~~~~~~~
\Gamma(\psi^\prime\to \psi(1^1P_1)+\pi^0)=0.17
\left(\frac{\alpha_M}{\alpha_E}\right)~{\rm keV},\nonumber\\
&&~~~~~~~~~~~~~~~~~~~~~~B(\psi^\prime\to\psi(1^1P_1)+\pi^0)=(6.1\pm 0.7)
\left(\frac{\alpha_M}{\alpha_E}\right)\times 10^{-4},
\label{result0}
\end{eqnarray}
\begin{eqnarray}                  
&&\theta=-12^\circ:~~~~~~~
\Gamma(\psi^\prime\to \psi(1^1P_1)+\pi^0)=0.06
\left(\frac{\alpha_M}{\alpha_E}\right)~{\rm keV},\nonumber\\
&&~~~~~~~~~~~~~~~~~~~~~~B(\psi^\prime\to\psi(1^1P_1)+\pi^0)=(2.2\pm 0.2)
\left(\frac{\alpha_M}{\alpha_E}\right)\times 10^{-4}.
\label{result-12}
\end{eqnarray}
Comparing the results in Eqs.(\ref{result0}) with those in Eqs.(\ref{KTY}), we
see that the results in the two models are close to each other. From
Eqs.(\ref{result0}) and (\ref{result-12}), we see that the state-mixing effect 
significantly reduces the rate due to the fact that the sign of 
$(f^{110}_{1211}+f^{201}_{1211})$ is opposite to that of 
$(f^{110}_{2011}+f^{001}_{2011})$.

The detection of the process (\ref{1P1prod}) depends on the capability
of the photon detector. In the two-body decay (\ref{1P1prod}), the
energies of $\psi(1^1P_1)$ and $\pi^0$ are fixed, i.e.,
$E_{\psi(1^1P_1)}=(M^2_{\psi^\prime}+M^2_{\psi(1^1P_1)}-m^2_{\pi^0})/
(2M_{\psi^\prime})$, $E_{\pi^0}=(M^2_{\psi^\prime}-M^2_{\psi(1^1P_1)}
+m^2_{\pi^0})/(2M_{\psi^\prime})$. $\pi^0$ decays $99\%$ into two photons.
The two photons in $\psi^\prime\to\psi(1^1P_1)+\gamma+\gamma$ with 
invariant mass $M^2_{\gamma\gamma}=m^2_{\pi^0}$ comes
mainly from the $\pi^0$ decay in (\ref{1P1prod})~\footnote{The branching
ratio of the process $\psi^\prime\to J/\psi+\pi^0$ is $(9.7\pm
2.1)\times 10^{-4}$ \cite{PDG}, so that this process can also be
detected. However, it can be clearly distinguished from the signal
process (\ref{1P1prod}) by the measured value of $\omega_1+\omega_2=E_{\pi^0}$
since $M_{J/\psi}$ and $M_{\psi(1^1P_1)}$ are quite different.}
since the branching ratios of 
the cascade electromagnetic transitions $\psi^\prime\to\eta_c^\prime(3590)
+\gamma\to \psi(1^1P_1)+\gamma+\gamma$ and $\psi^\prime\to \psi(1^3P_2)
+\gamma\to \psi(1^1P_1)+\gamma+\gamma$ are of the order of $10^{-5}$
and $10^{-6}\--10^{-5}$ , respectively \cite{Porter}. If the momenta of the 
two photons can be measured with sufficient accuracy, one can look for the 
monotonic invariant mass $M_{\gamma\gamma}$ as the signal. Once the energies 
$\omega_1$ and $\omega_2$ of the two photons are measured, the mass 
$M_{\psi(1^1P_1)}$ can be determined from $\omega_1+\omega_2=
E_{\pi^0}=(M^2_{\psi^\prime}-M^2_{\psi(1^1P_1)}+m^2_{\pi^0})/
(2M_{\psi^\prime})$. 
Now BES has already accumulated $3\times 10^6~\psi^\prime$ events. According to
the branching ratio in (\ref{result-12}) and taking into account a
$10\%$ detection efficiency, there can be $10^2\--10^3$ events of 
$\psi^\prime\to\psi(1^1P_1)+\gamma+\gamma$. Unfortunately, the present BES 
photon detector is not efficient enough to do this kind of measurement. So one 
should take certain decay products of $\psi(1^1P_1)$ as the signal.

To calculate the branching ratios of various $\psi(1^1P_1)$ decay
channels, the crucial thing is to estimate the radiative decay rate
$\Gamma(\psi(1^1P_1)\to\eta_c\gamma)$ and the hadronic decay rate
$\Gamma(\psi(1^1P_1)\to {\rm light~hadrons})$. Since state-mixing effect is 
supposed to be not important for $\psi(1^1P_1)$, we can simply do the 
single-channel calculation. 
In a recent paper \cite{Maltoni}, the rate 
$\Gamma(\psi(1^1P_1)\to\eta_c\gamma)$ was estimated by using the spin symmetry
\begin{eqnarray}                      
\Gamma(\psi(1^1P_1)\to\eta_c\gamma)=\bigg(
\frac{E^{(1^1P1)}_\gamma}{E^{(1^3P_J)}_\gamma}\bigg)^3
\Gamma(\psi(1^3P_J)\to\eta_c\gamma)
\label{relation}
\end{eqnarray}
and the experimental data of $\Gamma(\psi(1^3P_J)\to\eta_c\gamma)$. The
obtained value is \cite{Maltoni}
\begin{eqnarray}                      
\Gamma(\psi(1^1P_1)\to\eta_c\gamma)=0.45~{\rm MeV}.
\label{1P1-eta_cgamma}
\end{eqnarray}    
For $\Gamma(\psi(1^1P_1)\to {\rm light~hadrons})$, we take the
following approximation as in Ref.\cite{KTY}: 
\begin{eqnarray}                      
\displaystyle
\frac{\Gamma(\psi(1^1P_1)\to {\rm light~hadrons})}{\Gamma(J/\psi\to
{\rm light~hadrons})}\approx
\frac{\Gamma(\psi(1^1P_1)\to 3g)}{\Gamma(J/\psi\to 3g)}.
\label{1P1/3S1}
\end{eqnarray}
The rate $\Gamma(J/\psi\to 3g)$ can be obtained from \cite{KTY}
\begin{eqnarray}                       
\Gamma(J/\psi\to 3g)=[1-(2+R)B(J/\psi\to e^+e^-)]\Gamma_{tot}(J/\psi),
\label{J/psi-3g}
\end{eqnarray}
where $R=N_c(4/9+1/9+1/9)=2$.  Taking the updated data
$\Gamma_{tot}(J/\psi)=87\pm 5$~keV, $B(J/\psi\to e^+e^-)=(5.93\pm
0.10)\%$ \cite{PDG}, we obtain
\begin{eqnarray}                  
\Gamma(J/\psi\to 3g)=(66\pm 4)~{\rm keV}.
\label{Jpsi-3gno}
\end{eqnarray}

We first take the conventional perturbative QCD (PQCD) approach to calculate 
the ratio in Eq.(\ref{1P1/3S1}), in which 
it is assumed that the long-distance effect is factorized into the wave 
function of the quarkonium (naive factorization). The leading order PQCD 
formula for the right-hand side of Eq.(\ref{1P1/3S1}) has 
been given in Refs.\cite{PQCD,Remiddi}. In the Chen-Kuang potential model, the 
ratio on the right-hand side of Eq.(\ref{1P1/3S1}) has been calculated in 
Ref.\cite{CY}, which is 0.515. Thus we obtain
\begin{eqnarray}                   
\Gamma(\psi(1^1P_1)\to 3g)=(45\pm 3)~{\rm keV}.
\label{1P1-3gno}
\end{eqnarray}
Then from Eqs.(\ref{1P1-eta_cgamma}), (\ref{1P1-3gno}) and some
$\psi(1^1P_1)$ decay modes with small decay rates given in
Ref.\cite{CY}, we obtain, in PQCD,
\begin{eqnarray}                     
{\rm PQCD}:~~~~~~~~~~~~~~&&\Gamma_{tot}(\psi(1^1P_1))=(0.51\pm 0.01)~{\rm MeV},
\nonumber\\
&&B(\psi(1^1P_1)\to\eta_c\gamma)\approx (88\pm 2)\%,\nonumber\\
&&B(\psi(1^1P_1)\to {\rm light~hadrons})\approx (8.8\pm 0.8)\%.
\label{1P1decay1}
\end{eqnarray}
We see that both $~\psi(1^1P_1)\to \eta_c\gamma~$ and $~\psi(1^1P_1)\to
{\rm light~hadrons}~$ are detectable, and the most promising mode is
the radiative transition $~\psi(1^1P_1)\to\eta_c\gamma$.

Next we take the nonrelativistic QCD (NRQCD) approach developed in recent 
years \cite{BBL}. NRQCD is a more sophisticated approach in which the naive
factorization assumption is avoided. It has been intensively studied in
recent years and has been applied to studying the hadronic decays of
heavy quarkonia \cite{BBL,HC,Maltoni,BEPSV}. The study is up to
next-to-leading order (NLO) which contains unknown matrix elements of
some operators\footnote{In a recent paper \cite{Maltoni}, the NLO value 
$~\Gamma(\psi(1^1P_1))\to {\rm light~hadrons}=(0.72\pm 0.32)$ MeV is obtained. 
It is larger than the leading order value [cf. Eq.(\ref{1P1-LH_NRQCD})] by 
$26\%$ which is within the uncertainty ($44\%$) of the NLO result, and may be 
neglected relative to the large theoretical uncertainty and the large 
experimental errors in the input data in this study. Furthermore, the
uncertainty in this NLO result is so large that, together with the
experimental errors in the input data, it can hardly make clear
predictions for the final results.}. Considering the large theoretical 
uncertainty and the large experimental errors in the input data in this
study, taking the leading order (LO) NRQCD result concerning only one
matrix element related to the wave function at the origin is already 
sufficient for the present purpose. The LO NRQCD result of
$~\Gamma(\psi(1^1P_1)\to {\rm light~hadrons}~$ is \cite{BBL}
\begin{eqnarray}                     
{\rm NRQCD}:~~~~~~&&\Gamma[\psi(1^1P_1)\to {\rm light~hadrons}]=(0.53\pm 0.08)
~{\rm MeV}.
\label{1P1-LH_NRQCD}
\end{eqnarray}
This is very different from the corresponding values in Eq.(\ref{1P1-3gno}) in 
the conventional PQCD approach. With the updated value of 
$~\psi(1^1P_1)\to\eta_c\gamma~$ [cf. Eq.(\ref{1P1-eta_cgamma})], the NRQCD
predicted total width of $\psi(1^1P_1)$ will be slightly larger than that 
given in Ref.\cite{BBL}, which is 
\begin{eqnarray}                    
\!\!\!\!\!\!\!\!\!\!\!\!\!\!\!\!{\rm NRQCD}:~~~~~~~~~~~~~~~~~~
\Gamma_{tot}(\psi(1^1P_1)=(1.1\pm 0.09)~{\rm MeV}.
\label{1P1width}
\end{eqnarray}
Then we have
\begin{eqnarray}                      
\!\!\!\!\!\!\!\!\!{\rm NRQCD}:~~~~~~~~~~
&&B(\psi(1^1P_1)\to \eta_c\gamma)=(41\pm 3)\%,\nonumber\\
&&B(\psi(1^1P_1)\to {\rm light~hadrons})=(48\pm 7)\%.
\label {1P1decay2}
\end{eqnarray}
We see that the predictions in [cf. Eqs.(\ref{1P1decay2})] are quite
different from those of PQCD [cf. Eqs.(\ref{1P1decay1})]. Again, both 
$~\psi(1^1P_1)\to\eta_c\gamma~$ and $~\psi(1^1P_1)\to {\rm light~hadrons}~$ 
modes are detectable, but here $~\psi(1^1P_1)\to {\rm light~hadrons}~$ is 
slightly more promising.

Let us first consider the mode $\psi(1^1P_1)\to\eta_c\gamma$.
We then need to take certain detectable decay channels of $\eta_c$ as the 
signal. Possible modes are $\eta_c\to K\bar{K}\pi$, $\eta_c\to \rho\rho$, and 
$\eta_c\to \pi^+\pi^-K^+K^-$. The branching ratios are \cite{PDG}
\begin{eqnarray}                          
&&B(\eta_c\to K\bar{K}\pi)=(5.5\pm 1.7)\%,\nonumber\\
&&B(\eta_c\to \rho\rho)=(2.6\pm 0.9)\%,\nonumber\\
&&B(\eta_c\to \pi^+\pi^-K^+K^-)=(2.0^{+0.7}_{-0.6})\%.
\label{etacdecay}
\end{eqnarray}
Thus we can look for the decay chains
\begin{eqnarray}                          
&&\psi^\prime\to\psi(1^1P_1)\pi^0\to\psi(1^1P_1)\gamma\gamma\to
\eta_c\gamma\gamma\gamma\to K\bar{K}\pi\gamma\gamma\gamma,\nonumber\\
&&\psi^\prime\to\psi(1^1P_1)\pi^0\to\psi(1^1P_1)\gamma\gamma\to
\eta_c\gamma\gamma\gamma\to \rho\rho\gamma\gamma\gamma,\nonumber\\
&&\psi^\prime\to\psi(1^1P_1)\pi^0\to\psi(1^1P_1)\gamma\gamma\to
\eta_c\gamma\gamma\gamma\to K^+K^-\pi^+\pi^-\gamma\gamma\gamma.
\label{detect1}
\end{eqnarray}

If BES can accumulate $3\times 10^7$ $\psi^\prime$ events in the near
future, then from Eqs.(\ref{result-12}), (\ref{1P1decay1}) and 
(\ref{etacdecay}), and taking into account a $10\%$ detection efficiency, we 
obtain the event numbers of the signals in (\ref{detect1}) in the Chen-Kuang 
potential model for $\theta=-12^\circ$. The results are listed in Table I. The
corresponding event numbers in the NRQCD approach from Eqs.(\ref{result-12}), 
(\ref{1P1decay2}), and (\ref{etacdecay}) are listed in 
Table II. We see that the signals are detectable.

Next we consider the mode $~\psi(1^1P_1)\to {\rm light~hadrons}$. We
need to look at $\psi(1^1P_1)$ decaying into a certain exclusive hadronic
channel $h$. As in Ref.\cite{KTY,CY}, we assume that
\begin{eqnarray}                       
\displaystyle
\frac{\Gamma(\psi(1^1P_1)\to h)}{\Gamma(\eta_c\to h)}
\approx \frac{\Gamma(\psi(1^1P_1)\to 3g)}{\Gamma(\eta_c\to 2g)},
\label{1p1-h/etac-h}
\end{eqnarray}
and we can estimate $~\eta_c\to h~$ from Eq.(\ref{Jpsi-3gno}).

First, in the conventional PQCD, we have \cite{PQCD}
\begin{eqnarray}                       
\displaystyle
\frac{\Gamma(\eta_c\to 2g)}{\Gamma(J/\psi\to 3g)}
=\frac{27\pi}{5(\pi^2-9)\alpha_s}\bigg(\displaystyle\frac{ M^2_{J/\psi}}
{M^2_{\eta_c}}\bigg),
\label{etac/Jpsi}
\end{eqnarray}
which is model-independent. Then from Eq.(\ref{Jpsi-3gno}) we have
\begin{eqnarray}                         
\Gamma(\eta_c\to 2g)=(6.4\pm 0.4)~{\rm MeV},
\label{etac-2g}
\end{eqnarray}
in which we have taken $\alpha_s(M_{\eta_c})\approx 0.22$,
$M_{J/\psi}=3096.87\pm 0.04$ MeV, $M_{\eta_c}=2978.8\pm 1.8$ MeV \cite{PDG}.
So, with Eq.(\ref{1P1-3gno}), we have
\begin{eqnarray}                       
{\rm PQCD}:~~~~~~\Gamma(\psi(1^1P_1)\to h)=(0.010\pm 0.001)
\Gamma(\eta_c\to h).
\label{1P1-h_PQCD}
\end{eqnarray}

Next, for NRQCD. to LO, the ratio $~\Gamma(\eta_c\to 2g)/\Gamma(J/\psi\to 3g)~$
is the same as Eq.(\ref{etac/Jpsi}) \cite{BBL}. So that we still have 
Eq.(\ref{etac-2g}). Then from Eqs.(\ref{etac-2g}) and (\ref{1P1-LH_NRQCD}), we 
have 
\begin{eqnarray}                    
{\rm NRQCD}:~~~~~~\Gamma(\psi(1^1P_1)\to h)=(0.083\pm 0.018)
\Gamma(\eta_c\to h).
\label{1P1-h_NRQCD}
\end{eqnarray}

With the experimental value $\Gamma_{tot}(\eta_c)=13.2^{+3.8}_{-3.2}$ MeV
\cite{PDG}, we have
\begin{eqnarray}                       
{\rm PQCD}:~~~~~~~~
&&\Gamma(\psi(1^1P_1)\to K\bar{K}\pi)=(7.3\pm 3.0)~{\rm keV},\nonumber\\
&&\Gamma(\psi(1^1P_1)\to \rho\rho)=(3.4\pm 1.5)~{\rm keV},\nonumber\\
&&\Gamma(\psi(1^1P_1)\to \pi^+\pi^- K^+K^-)=(2.6\pm 1.1)~{\rm keV},\\
{\rm NRQCD}:~~~~~~
&&\Gamma(\psi(1^1P_1)\to K\bar{K}\pi)=(60\pm 32)~{\rm keV},\nonumber\\
&&\Gamma(\psi(1^1P_1)\to \rho\rho)=(28\pm 16)~{\rm keV},\nonumber\\
&&\Gamma(\psi(1^1P_1)\to \pi^+\pi^- K^+K^-)=(22\pm 12)~{\rm keV}.
\label{Gamma(1P1-h)}
\end{eqnarray}

Then with Eqs.(\ref{1P1decay1}) and (\ref{1P1-LH_NRQCD}) we have
\begin{eqnarray}                      
{\rm PQCD}:~~~~~~~~&&B(\psi(1^1P_1)\to K\bar{K}\pi)=(1.4\pm 0.9)\%,\nonumber\\
&&B(\psi(1^1P_1)\to \rho\rho)=(0.67\pm 0.43)\%,\nonumber\\
&&B(\psi(1^1P_1)\to \pi^+\pi^-K^+K^-)=(0.51\pm 0.32)\%,\\
{\rm NRQCD}:~~~~~~&&B(\psi(1^1P_1)\to K\bar{K}\pi)=(5.5\pm 3.3)\%,\nonumber\\
&&B(\psi(1^1P_1)\to \rho\rho)=(2.6\pm 1.7)\%,\nonumber\\
&&B(\psi(1^1P_1)\to \pi^+\pi^-K^+K^-)=(2.0\pm 1.2)\%.
\label{B(1P1-h)}
\end{eqnarray}
Now, instead of looking for the chains in (\ref{detect1}), one can look
for
\begin{eqnarray}                      
&&\psi^\prime\to\psi(1^1P_1)\pi^0\to\psi(1^1P_1)\gamma\gamma\to
K\bar{K}\pi\gamma\gamma,\nonumber\\
&&\psi^\prime\to\psi(1^1P_1)\pi^0\to\psi(1^1P_1)\gamma\gamma\to
\rho\rho\gamma\gamma,\nonumber\\
&&\psi^\prime\to\psi(1^1P_1)\pi^0\to\psi(1^1P_1)\gamma\gamma\to
K^+K^-\pi^+\pi^-\gamma\gamma.
\label{detect2}
\end{eqnarray}

Taking into account a $10\%$ detection efficiency, the numbers of events for 
the chains in Eqs.(\ref{detect2}) with $3\times 10^7$ $\psi^\prime$ events in 
the Chen-Kuang potential model for $\theta=-12^\circ$ are listed in
Table III (PQCD) and Table IV (NRQCD). We see that the two decay chains are 
all detectable, and the chains (\ref{detect1}) are more promising
in the PQCD approach, while the chains (\ref{detect2}) are more
promising in the NRQCD approach.

Comparing the numbers in Table I$\--$ Table IV, we see
that the NRQCD approach reduces the numbers of events for the chains
(\ref{detect1}) by roughly a factor of 2, and increases the numbers
of events for the chains (\ref{detect2}) by roughly a factor of 4
relative to the conventional PQCD approach.

The above situation shows that the detection of the chains
(\ref{detect1}) and (\ref{detect2}) is interesting not only for
the $\psi(1^1P_1)$ searches at BEPC, but also
for providing a test of the difference between the NRQCD predictions
and the conventional PQCD predictions.

In conclusion, our calculation shows that the $S-D$ mixing effect does
affect the transition rate (\ref{1P1prod}) significantly. It reduces
the rate (branching ration) by about a factor of 3 relative to the
single-channel result [cf. Eqs.(\ref{result0}) and (\ref{result-12})].
If the photon detector in the future BES III can be efficient enough
to measure the momenta of the two photons from $\pi^0\to \gamma\gamma$,
the search for the $\psi(1^1P_1)$ state via the process (\ref{1P1prod})
will not be difficult at BEPC II. At the present  BES,
one should tag $\psi(1^1P_1)$ directly via its decay products. Our
results in Table I$\--$Table IV show that the search for $\psi(1^1P_1)$
is possible with the present BES if an accumulation of $3\times 10^7$ 
$\psi^\prime$ events can be achieved in the near future.

Finally, we would like to mention that we have also calculated the rate of 
$\psi(3770)\to J/\psi~\pi\pi$ in
the Chen-Kuang potential model for $\theta=-12^\circ$ and
$\Lambda_{\overline{MS}}=200$ MeV. The result is $\Gamma(\psi(3770)\to
J/\psi~\pi\pi)=170~{\rm keV}~({\rm for}~c_2=3c_1),~{\rm or}~37~{\rm
keV}~({\rm for}~c_2=c_1)$. This is close to the corresponding values 
$\Gamma(\psi(3770)\to J/\psi~\pi\pi)=160~{\rm keV}~({\rm for}~c_2=3c_1),~{\rm 
or}~30~{\rm keV}~({\rm for}~c_2=c_1)$ in the Cornell potential model 
\cite{KY90}. Together with the results in Eqs.(\ref{result0}) and (\ref{KTY}), 
we see that the results are insensitive to the potential models used in
the calculation as compared with the uncertainties in the approach.

I would like to thank Hui Li and Zhi-Tong Yang for participating the 
calculation.
This work is supported by the National Natural Science Foundation of
China, the Foundation of Fundamental Research of Tsinghua University,
and a grant from BEPC National Laboratory.

\newpage
\begin{center}
{\large\bf Tables}
\end{center}
\begin{table}[t]
\null\noindent
TABLE I. Numbers of events for the processes in (\ref{detect1}) with
$3\times 10^7$ $\psi^\prime$ events in the Chen-Kuang potential
model for $\theta=-12^\circ$ and $\alpha_M/\alpha_E$=$1\--3$ with the 
conventional PQCD approach to $\Gamma[\psi(1^1P_1)\to {\rm light~hadrons}]$. 
A $10\%$ detection efficiency has been taken into account.
\begin{tabular}{cccc}
 &$\psi^\prime\to K\bar{K}\pi\gamma\gamma\gamma$&
 $\psi^\prime\to\rho\rho\gamma\gamma\gamma$
 &$\psi^\prime\to K^+K^=\pi^+\pi^=\gamma\gamma\gamma$\\
\hline
$\alpha_M=\alpha_E$&$32\pm 13$&$15\pm 7$&$12\pm 5$\\
$\alpha_M=2\alpha_E$&$64\pm 27$&$30\pm 14$&$24\pm 10$\\
$\alpha_M=3\alpha_E$&$96\pm 40$&$45\pm 21$&$35\pm 15$\\
\end{tabular}
\end{table}

\begin{table}[t]
\null\noindent
TABLE II Numbers of events for the processes in (\ref{detect1}) with
$3\times 10^7$ $\psi^\prime$ events in the Chen-Kuang potential model for 
$\theta=-12^\circ$ and $\alpha_M/\alpha_E$=$1\--3$ with the NRQCD approach to 
$\Gamma[\psi(1^1P_1)\to {\rm light~hadrons}]$. 
A $10\%$ detection efficiency has been taken into account.
\begin{tabular}{cccc}
 &$\psi^\prime\to K\bar{K}\pi\gamma\gamma\gamma$&
 $\psi^\prime\to\rho\rho\gamma\gamma\gamma$
 &$\psi^\prime\to K^+K^=\pi^+\pi^=\gamma\gamma\gamma$\\
\hline
$\alpha_M=\alpha_E$&$15\pm 7$&$7\pm 4$&$5\pm 3$\\
$\alpha_M=2\alpha_E$&$30\pm 22$&$14\pm 7$&$11\pm 5$\\
$\alpha_M=3\alpha_E$&$45\pm 21$&$21\pm 11$&$16\pm 8$\\
\end{tabular}
\end{table}

\begin{table}[h]
\null\noindent
TABLE III Numbers of events for the processes in (\ref{detect2}) with
$3\times 10^7$ $\psi^\prime$ events in the Chen-Kuang potential
model for $\theta=-12^\circ$ and $\alpha_M/\alpha_E$=$1\--3$ with the
conventional PQCD approach to $\Gamma[\psi(1^1`P_1)\to {\rm light~hadrons}]$. 
A $10\%$ detection efficiency has been taken into account.
\begin{tabular}{cccc}
 &$\psi^\prime\to K\bar{K}\pi\gamma\gamma$&
 $\psi^\prime\to\rho\rho\gamma\gamma$
 &$\psi^\prime\to K^+K^=\pi^+\pi^=\gamma\gamma$\\
\hline
$\alpha_M=\alpha_E$&$9\pm 7$&$4\pm 3$&$3\pm 2$\\
$\alpha_M=2\alpha_E$&$18\pm 13$&$9 \pm 6$&$7\pm 5$\\
$\alpha_M=3\alpha_E$&$28\pm 20$&$13\pm 10$&$10\pm 7$\\
\end{tabular}
\end{table}

\begin{table}[h]
\null\noindent
TABLE IV Numbers of events for the processes in (\ref{detect2}) with
$3\times 10^7$ $\psi^\prime$ events in the Chen-Kuang potential
model for $\theta=-12^\circ$ and $\alpha_M/\alpha_E$=$1\--3$ with the
NRQCD approach to $\Gamma[\psi(1^1`P_1)\to {\rm light~hadrons}]$. 
A $10\%$ detection efficiency has been taken into account.
\begin{tabular}{cccc}
 &$\psi^\prime\to K\bar{K}\pi\gamma\gamma$&
 $\psi^\prime\to\rho\rho\gamma\gamma$
 &$\psi^\prime\to K^+K^=\pi^+\pi^=\gamma\gamma$\\
\hline
$\alpha_M=\alpha_E$&$36\pm 25$&$17\pm 13$&$13\pm 9$\\
$\alpha_M=2\alpha_E$&$73\pm 51$&$34 \pm 25$&$26\pm 19$\\
$\alpha_M=3\alpha_E$&$109\pm 76$&$52\pm 41$&$40\pm 28$\\
\end{tabular}
\end{table}

\end{document}